\documentclass[%
 aip,
 sd,%
 amsmath,amssymb,
 reprint,%
]{revtex4-1}

\usepackage{graphicx}
\usepackage{dcolumn}
\usepackage{bm}

\begin{document}

\preprint{AIP/123-QED}

\title { Out-of-plane easy-axis in thin films of diluted magnetic semiconductor Ba$_{1-x}$K$_x$(Zn$_{1-y}$Mn$_y$)$_2$As$_2$ }


\author{R. Wang}
\affiliation{ Beijing National Laboratory for Condensed Matter Physics, Institute of Physics, University of Chinese Academy of Sciences, Chinese Academy of Sciences, Beijing 100190, China }

\author{Z. X. Huang}
\affiliation{ Beijing National Laboratory for Condensed Matter Physics, Institute of Physics, University of Chinese Academy of Sciences, Chinese Academy of Sciences, Beijing 100190, China }

\author{G. Q. Zhao}
\affiliation{ Beijing National Laboratory for Condensed Matter Physics, Institute of Physics, University of Chinese Academy of Sciences, Chinese Academy of Sciences, Beijing 100190, China }

\author{S. Yu}
\affiliation{ Beijing National Laboratory for Condensed Matter Physics, Institute of Physics, University of Chinese Academy of Sciences, Chinese Academy of Sciences, Beijing 100190, China }

\author{Z. Deng}
\affiliation{ Beijing National Laboratory for Condensed Matter Physics, Institute of Physics, University of Chinese Academy of Sciences, Chinese Academy of Sciences, Beijing 100190, China }

\author{C. Q. Jin}
\affiliation{ Beijing National Laboratory for Condensed Matter Physics, Institute of Physics, University of Chinese Academy of Sciences, Chinese Academy of Sciences, Beijing 100190, China }

\author{Q. J. Jia}
\affiliation{ Institute of High Energy Physics, Chinese Academy of Sciences, Beijing 100049, China }

\author{Y. Chen}
\affiliation{ Institute of High Energy Physics, Chinese Academy of Sciences, Beijing 100049, China }

\author{T. Y. Yang}
\affiliation{ Shanghai Institute of Applied Physics, Chinese Academy of Sciences, Shanghai 201204, China }

\author{X. M. Jiang}
\affiliation{ Institute of High Energy Physics, Chinese Academy of Sciences, Beijing 100049, China }
\affiliation{ Beijing Advanced Sciences and Innovation Center, Chinese Academy of Sciences, Beijing 101407, China }

\author{L. X. Cao}
\email{Author to whom correspondence should be addressed. lxcao@iphy.ac.cn }
\affiliation{ Beijing National Laboratory for Condensed Matter Physics, Institute of Physics, University of Chinese Academy of Sciences, Chinese Academy of Sciences, Beijing 100190, China }

\date{\today}

\begin{abstract}

Single-phased, single-oriented thin films of Mn-doped ZnAs-based diluted magnetic semiconductor (DMS) Ba$_{1-x}$K$_x$(Zn$_{1-y}$Mn$_y$)$_2$As$_2$ (x = 0.03, 0.08; y = 0.15) have been deposited on Si, SrTiO$_3$, LaAlO$_3$, (La,Sr)(Al,Ta)O$_3$, and MgAl$_2$O$_4$ substrates, respectively.
Utilizing a combined synthesis and characterization system excluding the air and further optimizing the deposition parameters, high-quality thin films could be obtained and be measured showing that they can keep inactive-in-air up to more than 90 hours characterized by electrical transport measurements.
In comparison with films of x = 0.03 which possess relatively higher resistivity, weaker magnetic performances, and larger energy gap, thin films of x = 0.08 show better electrical and magnetic performances.
Strong magnetic anisotropy was found in films of x = 0.08 grown on (La,Sr)(Al,Ta)O$_3$ substrate with their magnetic polarization aligned almost solely on the film growth direction.

\end{abstract}

\maketitle

Diluted magnetic semiconductor (DMS) thin films \cite{ohno96APL,ohno98Science,ohno14RMP} attracted much attentions in the past several decades since semiconductors possessing intrinsic ferromagnetic properties not only can integrate different functionalities, such as electric, magnetic, and optic properties, into one device; but also may lead to new findings in physics as well. \cite{ohno14RMP}
However, some issues have to be tackled and solved before the real applications of DMS materials, especially Curie temperature T$_c$ which is still far below the room temperature.
In this respect, as a result of 20 years heavy investigation, record T$_c$ of (Ga,Mn)As film has been boosted from the initial 75 K \cite{ohno96APL} to $\sim$185 K until now. \cite{wang08}
Very recently Mn-doped ZnAs-based DMS materials \cite{2011Jin111,2013Jin122} emerged, providing an alternative to study on new DMS thin films possessing better properties, especially under the circumstances that preliminary evidences in the bulk reported possible T$_c$ up to 230 K \cite{JinCSB2014}.

In this paper we deposited and studied the properties of the Ba$_{1-x}$K$_x$(Zn$_{1-y}$Mn$_y$)$_2$As$_2$ (BKZMA) (x = 0.03, 0.08; y = 0.15) DMS $\bf{films}$ for the first time.
Over 320 films have been studied.
Since the new ZnAs-based materials in polycrystal form are metastable and deteriorate slowly in air, \cite{DengZthesis} we developed a combined system which enables the whole process of pulsed laser deposition (PLD), X-ray diffraction (XRD), electrode metal deposition, ultrasonic wire bonding, and low-temperature electrical transport measurement all performed in an environment protected which excludes the air.
The system is fulfilled with 1 atm nitrogen with trace water and oxygen contaminations maintained incessantly less than 1$\times$10$^{-7}$ each.
The PLD chamber can be pumped down from 1 atm to 8.5$\times$10$^{-7}$ Pa.
Sintered bulk composite of 15 mm in diameter and with nominal formula Ba$_{0.7}$K$_{0.4}$(Zn$_{0.85}$Mn$_{0.15}$)$_2$As$_2$ was used as target,
which was irradiated by a scanned KrF excimer laser beam with power, repetition rate, and focused beam size of 240 mJ, 2 Hz, and $\sim$3 mm$^2$, respectively.
The films were deposited at around 540 $^{\circ}$C on (001)-oriented Si, SrTiO$_3$ (STO), LaAlO$_3$ (LAO), (La,Sr)(Al,Ta)O$_3$ (LSAT), and MgAl$_2$O$_4$ (MAO) substrates, respectively.
After pumping down to the background pressure of $\sim$1$\times$10$^{-6}$ Pa, during deposition the chamber was either filled in with argon gas of $\sim$5 Pa or was kept pumping.
The distance between the target and substrate (d) was from 25 mm to 35 mm.
Besides characterized by $\it{in}$-$\it{situ}$ XRD and low-temperature electrical transport measurement, some films were also characterized by $\it{ex}$-$\it{situ}$ XRD, energy dispersive analysis of X-ray (EDAX), low-temperature magnetic measurement, and X-ray reflectivity (XRR) measurement performed at BL14B1 station of Shanghai Synchrotron Radiation Facility (SSRF) and 1W1A station of Beijing Synchrotron Radiation Facility (BSRF).

Study on BKZMA films faces 2 main technical problems,
(1) relatively low Potassium composition could be reached in the deposited films, compared to their bulk target counterparts, because of low melting point and thus high gas pressure of Potassium;
(2) short lifetimes of the films.
The DMS bulk materials including BKZMA deteriorate slowly in air \cite{DengZthesis}, and the polycrystalline films with large surface survive even much shorter than their bulk counterparts.
In the very beginning of our study, although the deposited films were observed by naked eyes homogeneous and shinning grey when they were in the protected environment excluding the air, however, such shinning grey surfaces evanesced very quickly in a time scale of around 1-5 seconds when taking the film samples out of the protected environment.
Such films were found not single-oriented although they were single-phased, measured by our $\it{in}$-$\it{situ}$ XRD as shown in Fig. 1(a) for one example.
The integrated intensities counts of the (103)- and (004)- diffraction peaks are 826 and 7720, respectively; corresponding to 0.8$\%$ and 99.2$\%$ volume fractions of (103)- and (001)- oriented grains in the films, respectively. \cite{XRDpdfCard}
The reasons of the short lifetime of the films might be that grain boundaries among different orientated grains might help quicker diffusions into the film by the water and oxygen molecules which react with the BKZMA grains.
By carefully optimizing the deposition parameters as will be given below, the shinning surfaces did not change after exposing the films in air for more than 7 days.
The key parameter responsible for eliminating the minor volume fraction of (103)-oriented grains is deposition temperature, which was optimized at 540 $^{\circ}$C finally.

Furthermore, we noticed that deposition pressure (P) and target-substrate distance (d) determine the Potassium composition of the films, i.e., the higher the pressure and the smaller the distance, the higher the Potassium composition; the lower the pressure and the larger the distance, the lower the Potassium composition.
Helped by the EDAX (results not shown), we found that Potassium composition x can be fixed at 0.08 when P = 5 Pa and d = 25 mm, while x = 0.03 when P $<$ 1$\times$10$^{-5}$ Pa (pumping during deposition) and d = 35 mm.
As some examples, films of x = 0.03 grown on LAST, STO, and LAO were measured not only by $\it{in}$-$\it{situ}$ but also by $\it{ex}$-$\it{situ}$ XRD, as shown in Fig. 1(b)-(d); while films of x = 0.08 grown on Si and MAO were measured by $\it{in}$-$\it{situ}$ XRD, as shown in Fig. 1(e)-(f).
The $\it{c}$-axis lattice constants of x = 0.03 films calculated from Fig. 1(b)-(d) are 13.46$\pm$0.01 {\AA}, while that of x = 0.08 films from Fig. 1(a) and Fig. 1(e)-(f) are 13.41$\pm$0.01 {\AA}; which means higher Potassium concentration leading to $\it{c}$-axis lattice shrinkage.
Note, the lattice constants of substrate single-crystals were used as inner standard when calculating the constants of films.

\begin{figure}
\caption{
XRD spectra of Ba$_{1-x}$K$_x$(Zn$_{0.85}$Mn$_{0.15}$)$_2$As$_2$ (x = 0.03, 0.08) films characterized
$\it{in}$-$\it{situ}$ or $\it{ex}$-$\it{situ}$, on substrates of (a) LSAT, $\it{in}$-$\it{situ}$;
(b) LSAT, $\it{ex}$-$\it{situ}$; (c) SrTiO$_3$, $\it{ex}$-$\it{situ}$; (d) LaAlO$_3$, $\it{ex}$-$\it{situ}$;
(e) Si, $\it{in}$-$\it{situ}$; and (f) MgAl$_2$O$_4$, $\it{in}$-$\it{situ}$.
Film shown in (a) is (001)-oriented mixed with minor amount of (103)-oriented grains, while others shown
in (b)-(f) are all (001)- single-oriented.
All films shown had been $\it{in}$-$\it{situ}$ XRD characterized.
Two vertical straight lines are guide to the eyes to indicate angle positions of (004) and (008) diffraction
peaks of x = 0.03 films.
Note the angle differences among peak positions of x = 0.03 and those of x = 0.08 films.
\label{}}%
\end{figure}

In order to further characterize the stability of such high-quality, single-oriented films, the resistance versus temperature of 5 randomly-selected films were measured after exposing them in air for every certain period of time, each period lasting from every 2 hours up to several 10 hours. The measured curves (not shown) were found almost the same up to $\sim$90 hours in comparison with the one measured initially without air exposure.

For electrical response of the films, low-temperature resistivities of the films with Potassium composition x = 0.03 and 0.08, grown on STO, LSAT, and LAO substrates, respectively, were measured $\it{in}$-$\it{situ}$, as shown in Fig. 2.
It is obvious that higher doping of Potassium leads to more carriers and therefore lower resistivity, roughly 1-2 orders of differences among 2 bunches of curves, c.f., Fig. 2.
The energy gaps fitted in the temperature range close to room temperature (from 300 K down to $\sim$150 K) via the thermal activation function \cite{NingEPL2014, NingSR2015}
$\rho \propto exp(E_g/2k_BT)$  are 50 eV, 42 eV, and 42 eV for x = 0.03 films on STO, LSAT, and LAO, respectively; as well as 27 eV, 32 eV, and 22 eV for x = 0.08 films correspondingly.
The results show that the more the introduction of the carriers into the films, the smaller the energy gap opening.
It should be noted that, (1) when temperature goes down from 300 K to 2 K, the bulk polycrystals show almost equal values in resistivity\cite{2013Jin122}, while our thin films show 2-3 orders of resistivity change;
(2) in comparison with the bulk polycrystals \cite{2013Jin122}, thin films possess resistivity $\sim$2 orders less at 300 K than those of bulk polycrystals; while their resistivities are almost the same at 2 K.
These may suggest that the new ZnAs-based DMS materials are highly anisotropic in electrical transport properties.

\begin{figure}
\caption{
Temperature dependencies of resistivities of Ba$_{1-x}$K$_x$(Zn$_{0.85}$Mn$_{0.15}$)$_2$As$_2$
(x = 0.03, 0.08) films grown on SrTiO$_3$, LSAT, and LaAlO$_3$ substrates, respectively.
\label{}}%
\end{figure}

As for magnetic performances of the $\bf{Diluted}$ magnetic semiconductor (DMS) $\bf{films}$ (the measured mass quantity is 6.3$\times$10$^{-6}$ grams for example, for a DMS film 100 nm thick and 4$\times$4 mm$^2$ large), it is a challenging task to characterize them precisely, considering magnetic contributions from the beneath $\sim$1 mm thick substrates.
The oxide dielectric crystals possess at least intrinsic diamagnetic responses as well as paramagnetic responses originating from tiny amount of impurities even down to ppm level. \cite{d0sub, d0Theo}
In Fig. 3(a)-(b), one x = 0.08 film grown on LSAT were measured under magnetic field of 500 Oe; while in Fig. (c), one LSAT substrate was measured, too. Correspondingly the signals originating solely from the film deduced from the above data are reproduced in Fig. 3(d)-(e).
It can be seen easily that, (1) the film is highly magnetic anisotropic with its magnetic polarization aligned almost entirely along $\it{c}$-axis of the film;
(2) the magnetic responses as function of temperature measured under field cooling (FC) and zero field cooling (ZFC) deviate from each other at $\sim$250 K, which may suggest a magnetic transition there; \cite{JinCSB2014}
(3) the M-T curve under ZFC along $\it{c}$-axis of the film (Fig. 3(d)) slightly goes down when temperature smaller than $\sim$160 K, which may be an indication of existence of spin glass \cite{NingSR2015}.
And therefore the film might be phase-seperated with spin glass as well as DMS phase all inside.
Our present study on the thin films is consistent with the previous one on the bulks. \cite{2013Jin122}
For example, as shown in Fig. 3B and Table 1 of Ref. 6, for a Ba$_{0.8}$K$_{0.2}$(Zn$_{0.9}$Mn$_{0.1}$)$_2$As$_2$ bulk sample, the Tc of which is 135 K, the ferromagnetic order is not fully developed in the entire volume of the bulk above 20 K up till 135 K.

Our experimental results reveal 2 important points of the study on new DMS materials:
(1) Both the electrical and magnetic transport measurements reveal the strong anisotropy in the films.
(2) Further studies should be focused not only on boosting the ferromagnetic transition temperature but also on enhancing the ferromagnetic orderings in the DMS.

\begin{figure}
\caption{
Temperature dependencies of magnetization measured under field cooling (FC) or zero field cooling (ZFC);
with the field perpendicular to or parallel to film plane, as well as to substrate surface.
(a), (b), measurement on x = 0.08 film grown on LSAT. (c) measurement on LSAT. (d), (e) signals of the films
deduced from (a)-(c).
\label{}}%
\end{figure}

The magnetization curves of the same film sample and also the LSAT substrate were measured at 2 K, as shown in Fig. 4.
It can be seen that saturation magnetization M$_s$, remanent magnetization M$_r$, and coercivity H$_c$ are 1.50 ${\mu}_B$/Mn, 0.15 ${\mu}_B$/Mn, and 400 Oe, respectively, measured with field perpendicular to the film; as well as 0.21 ${\mu}_B$/Mn, 0 ${\mu}_B$/Mn, and 0 Oe, respectively, with field parallel to the film.

\begin{figure}
\caption{
Magnetic hysteresis measurement on the film sample shown in Fig. 3, and magnetization measurement on LSAT
crystal shown in Fig. 3(c).
The inset gives the enlargement of the data.
\label{}}%
\end{figure}

In summary, we synthesized for the first time the $\bf{films}$ of new ZnAs-based DMS, Ba$_{1-x}$K$_x$(Zn$_{1-y}$Mn$_y$)$_2$As$_2$ (x = 0.03, 0.08; y = 0.15). The films grown on Si, SrTiO$_3$, LaAlO$_3$, (La,Sr)(Al,Ta)O$_3$, and MgAl$_2$O$_4$ substrates are not only single-phased but also single-oriented, showing strong electrical transport anisotropy as well as strong magnetic anisotropy with polarization aligned along $\it{c}$-axis of the film.
Further investigations on Hall measurements are under way trying to illustrate more physics of this new type of material.

\begin{acknowledgments}

This work was financially supported by the National Natural Science Foundation of China (11474324, 11534016, 11405253, and U1332205), as well as the Strategic Priority Research Program (XDB07020100) and the Youth Innovation Promotion Association (2016237) of the Chinese Academy of Sciences.

\end{acknowledgments}



\begin{thebibliography}{9}\label{sec:TeXbooks}


\bibitem{ohno96APL}
 H. Ohno, A. Shen, F. Matsukura, A. Oiwa, A. Endo, S. Katsumoto, and Y. Iye,
Appl. Phys. Lett. \textbf{69}, 363 (1996).

\bibitem{ohno98Science}
H. Ohno, Science \textbf{281}, 951 (1998).

\bibitem{ohno14RMP}
T. Dietl and H. Ohno,
Rev. Mod. Phys. \textbf{86}, 187 (2014).
\bibitem{wang08}
M. Wang, R. P. Campion, A. W. Rushforth, K. W. Edmonds, C. T. Foxon, and B. L. Gallagher,
Appl. Phys. Lett. \textbf{93}, 132103 (2008).

\bibitem{2011Jin111}
Z. Deng, C. Q. Jin, Q. Q. Liu, X.C. Wang, J. L. Zhu, S. M. Feng, L. C. Chen, R. C. Yu, C. Arguello, T. Goko, F. L. Ning, J. S. Zhang, Y. Y. Wang, A. A. Aczel, T. Munsie, T. J. Williams, G. M. Luke, T. Kakeshita, S. Uchida, W. Higemoto, T. U. Ito, B. Gu, S. Maekawa, G. D. Morris, and Y. J. Uemura, Nature Commun. \textbf{2}, 442 (2011).

\bibitem{2013Jin122}
K. Zhao, Z. Deng, X. C. Wang, W. Han, J. L. Zhu, X. Li, Q. Q. Liu, R. C. Yu, T. Goko, B. Frandsen, L. Liu, F. L. Ning, Y. J. Uemura, H. Dabkowska, G. M. Luke, H. Luetkens, E. Morenzoni, S. R. Dunsiger, A. Senyshyn, P. B{\"o}ni, and C. Q. Jin,
 Nature Commun. \textbf{4}, 1442 (2013).

\bibitem{JinCSB2014}
K. Zhao, B. J. Chen, G. Q. Zhao,
Z. Yuan, Q. Q.Liu, Z. Deng,
J. L. Zhu, and C. Q. Jin,
Chin. Sci. Bull. \textbf{59}, 2524 (2014).

\bibitem{DengZthesis}
 Z. Deng, Ph.D. thesis, University of Chinese Academy of Sciences, Beijing, 2012.


\bibitem{XRDpdfCard}
ICDD PDF-4 files 01-0765124 and 00-031-0154, from which the powder diffraction intensities of the (103) and (004) diffractions peaks are 100 and 7.7, respectively, for BaZn$_2$As$_2$; as well as 100 and 5, respectively, for BaMn$_2$As$_2$.



\bibitem{NingEPL2014}
C. Ding, Z. Gong, H. Y. Man, G. X. Zhi, S. L. Guo, Y. Zhao, H. D. Wang, B. Chen, and F. L. Ning,
 Europhys. Lett. \textbf{107}, 17004 (2014).

\bibitem{NingSR2015}
H. Y. Man, S. L. Guo, Y. Sui, Y. Guo, B. Chen, H. D. Wang, C. Ding, and F. L. Ning,
Sci. Rep. \textbf{5}, 15507 (2015).

\bibitem{d0sub}
M. Venkatesan, Pravin Kavle, S. B. Porter, K. Ackland, and J. M. D. Coey, IEEE Trans. Magn. \textbf{50}, 2201704 (2014).

\bibitem{d0Theo}
J. M. D. Coey, M. Venkatesan, and C. B. Fitzgerald, Nat. Mater. \textbf{4}, 173 (2005).







\end{thebibliography}
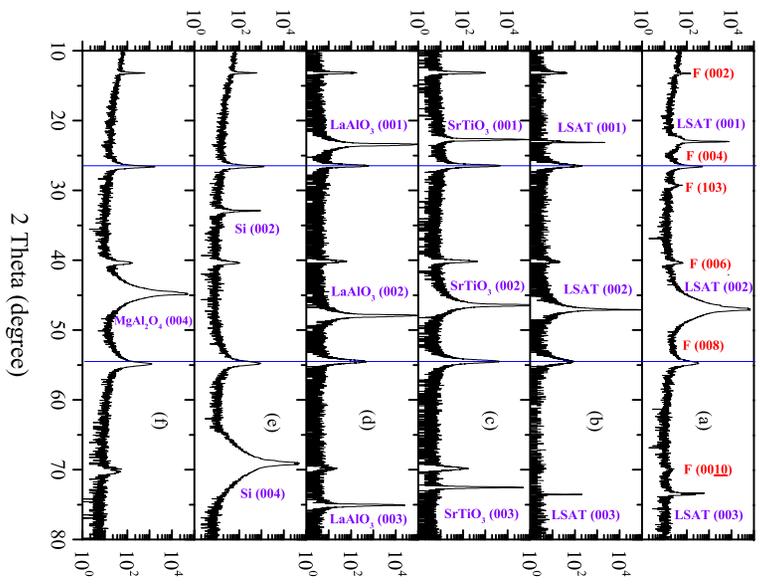
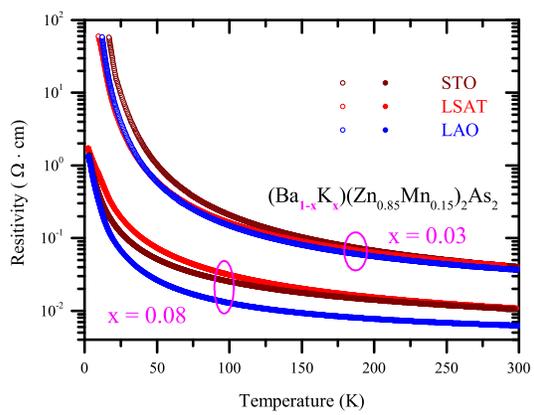
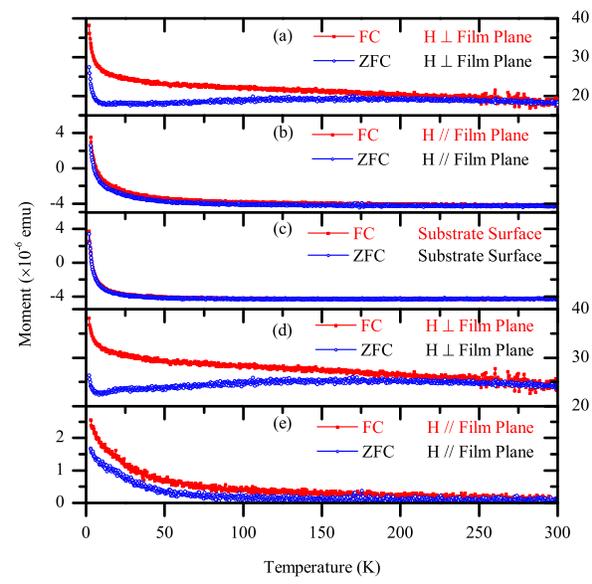
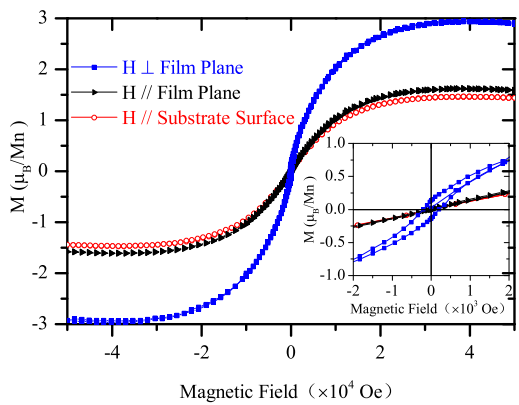
\end{document}